\documentclass[11pt]{article}
\usepackage[a4paper]{geometry}
\usepackage{amsfonts, amsmath, amssymb, amsthm, graphicx, caption, authblk, multirow, makecell, framed, float, tikz, xcolor, mathrsfs, changepage}
\usetikzlibrary{shapes,arrows,positioning}
\theoremstyle{definition}

\theoremstyle{remark}

\begin{document}
\tikzstyle{block} = [draw, rectangle]
\tikzstyle{line} = [draw, ->]
\tikzstyle{branch} = [draw, diamond]
\tikzstyle{finish} = [draw, circle]
\title{AND Protocols Using Only Uniform Shuffles}
\author[1]{Suthee Ruangwises\thanks{\texttt{ruangwises.s.aa@m.titech.ac.jp}}}
\author[1]{Toshiya Itoh\thanks{\texttt{titoh@c.titech.ac.jp}}}
\affil[1]{Department of Mathematical and Computing Science, Tokyo Institute of Technology, Tokyo, Japan}
\date{}
\maketitle

\begin{abstract}
Secure multi-party computation using a deck of playing cards has been a subject of research since the ``five-card trick'' introduced by den Boer in 1989. One of the main problems in card-based cryptography is to design \textit{committed-format} protocols to compute a Boolean AND operation subject to different runtime and shuffle restrictions by using as few cards as possible. In this paper, we introduce two AND protocols that use only \textit{uniform} shuffles. The first one requires four cards and is a \textit{restart-free} Las Vegas protocol with finite expected runtime. The second one requires five cards and always terminates in finite time.

\end{abstract}

\section{Introduction}
\subsection{The Five-Card Trick}
The concept of card-based cryptography started in 1989 with the ``five-card trick'' introduced by den Boer \cite{denboer}. In the original problem, Alice and Bob want to know whether they both like each other. However, no one wants to confess first because of fear of embarrassment if he/she gets rejected. Therefore, they need a protocol that only distinguishes the two cases where they both like each other and otherwise, without leaking any other information.

This situation is equivalent to Alice having a bit $a$ and Bob having a bit $b$ of either 0 or 1. Such protocol outputs the result of a Boolean operation $\text{AND}(a,b) = a \wedge b$ without leaking unnecessary information, i.e. if a player's bit is 1, he/she inevitably knows the other player's bit after knowing $a \wedge b$; if a player's bit is 0, he/she should know nothing about the other player's bit.

Following is the description of the five-card trick protocol, using three identical $\clubsuit$ cards and two identical $\heartsuit$ cards. Throughout this paper, we encode the bit 0 by the commitment $\clubsuit$$\heartsuit$ and 1 by the commitment $\heartsuit$$\clubsuit$. Initially, we give each player two cards, one $\clubsuit$ and one $\heartsuit$. We also have another $\clubsuit$ card faced down on the middle of a table. Alice places her two (face-down) cards encoding $a$ to the left of the middle card, while Bob places his two (face-down) cards encoding $b$ to the right of the middle card. There are following four possible sequences of the cards.

\begin{figure}[H]
    \centering
    \begin{minipage}{0.2\textwidth}
        \centering
        $\clubsuit$$\heartsuit$$\clubsuit$$\clubsuit$$\heartsuit$ \\
        $a=0$, $b=0$
    \end{minipage}
    \begin{minipage}{0.2\textwidth}
        \centering
        $\clubsuit$$\heartsuit$$\clubsuit$$\heartsuit$$\clubsuit$ \\
        $a=0$, $b=1$
    \end{minipage}
    \begin{minipage}{0.2\textwidth}
        \centering
        $\heartsuit$$\clubsuit$$\clubsuit$$\clubsuit$$\heartsuit$ \\
        $a=1$, $b=0$
    \end{minipage}
    \begin{minipage}{0.2\textwidth}
        \centering
        $\heartsuit$$\clubsuit$$\clubsuit$$\heartsuit$$\clubsuit$ \\
        $a=1$, $b=1$
    \end{minipage}
\end{figure}

Then, we swap the fourth and the fifth card, resulting in the following four possible sequences.

\begin{figure}[H]
    \centering
    \begin{minipage}{0.2\textwidth}
        \centering
        $\clubsuit$$\heartsuit$$\clubsuit$$\clubsuit$$\heartsuit$ \\
        $\Downarrow$ \\
        $\clubsuit$$\heartsuit$$\clubsuit$$\heartsuit$$\clubsuit$ \\
        $a=0$, $b=0$
    \end{minipage}
    \begin{minipage}{0.2\textwidth}
        \centering
        $\clubsuit$$\heartsuit$$\clubsuit$$\heartsuit$$\clubsuit$ \\
        $\Downarrow$ \\
        $\clubsuit$$\heartsuit$$\clubsuit$$\clubsuit$$\heartsuit$ \\
        $a=0$, $b=1$
    \end{minipage}
    \begin{minipage}{0.2\textwidth}
        \centering
        $\heartsuit$$\clubsuit$$\clubsuit$$\clubsuit$$\heartsuit$ \\
        $\Downarrow$ \\
        $\heartsuit$$\clubsuit$$\clubsuit$$\heartsuit$$\clubsuit$ \\
        $a=1$, $b=0$
    \end{minipage}
    \begin{minipage}{0.2\textwidth}
        \centering
        $\heartsuit$$\clubsuit$$\clubsuit$$\heartsuit$$\clubsuit$ \\
        $\Downarrow$ \\
        $\heartsuit$$\clubsuit$$\clubsuit$$\clubsuit$$\heartsuit$ \\
        $a=1$, $b=1$
    \end{minipage}
\end{figure}

Observe that there are only two possible sequences in a cyclic rotation of the deck, and the two $\heartsuit$ cards are adjacent to each other in the cycle only in the case that $a=1$ and $b=1$ (while all other three cases result in another same sequence), hence we can determine whether $a \wedge b = 1$ by looking at the cycle. We can obscure the initial position of the cards by shuffling the deck into a uniformly random cyclic permutation, i.e. a permutation uniformly chosen from $\{\text{id}, (1 2 3 4 5), (1 2 3 4 5)^2, (1 2 3 4 5)^3, (1 2 3 4 5)^4\}$ at random.

Mizuki et al. \cite{mizuki12} later improved the five-card trick protocol so that it requires only four cards instead of five. While both protocols are useful, the format of the output value $a \wedge b$ is different from the format of the inputs $a$ and $b$ ($\clubsuit$$\heartsuit$ for 0 and $\heartsuit$$\clubsuit$ for 1). Both protocols have drawback in the case that we want to compute an AND operation over three or more inputs. If a protocol is \textit{committed-format}, i.e. the output is encoded in the same format as the input, we can perform that protocol on an AND operation over the first two inputs, and use the output as an input of another AND operation with the third input, then with the fourth input, and so on. Therefore, most studies so far have been focused only on committed-format protocols.

\subsection{Properties of Protocols}
In the formal computation model of card-based protocols developed by Mizuki and Shizuya \cite{mizuki14}, a shuffle of the deck is mathematically defined by a pair ($\Pi$, $\mathscr{F}$), where $\Pi$ is a set of permutations and $\mathscr{F}$ is a probability distribution on $\Pi$. We call the shuffle \textit{uniform} if $\mathscr{F}$ is a uniform distribution, and \textit{closed} if $\Pi$ is a subgroup (of the symmetric group) \cite{abe}. Uniformness and closedness have practical benefits. A closed shuffle can be securely performed by letting the first player rearrange the deck into his selected permutation from $\Pi$ without the second player observing, then the second player do the same without the first player observing. Closedness guarantees that performing the shuffle twice still results in a permutation in $\Pi$, while uniformness makes it easier and more natural for a player to randomly select a permutation from $\Pi$.

In term of runtime, a protocol is called \textit{finite} if it is guaranteed to terminate after a finite number of steps. Apart from finite protocols, many studies have been focused on other protocols that are Las Vegas with finite expected runtime and \textit{restart-free}, i.e. players are required to put their commitments to the deck only once, not having to restart the whole process again.

\subsection{Previous Protocols}
In 1993, Cr\'{e}peau and Kilian \cite{crepeau} developed the first committed-format AND protocol using ten cards with four colors. Niemi and Renvall \cite{niemi} also developed another protocol using 12 cards but with only two colors. Stiglic \cite{stiglic} later reduced the number of required cards to eight. More recently in 2009, Mizuki and Sone \cite{mizuki09} developed an AND protocol using only six cards. This was an important milestone since their protocol was the first one that has finite runtime.

Koch et al. \cite{koch} investigated a novel way of shuffles that are not uniform or closed. That reduced the number of cards to five for finite protocol, and four for Las Vegas protocol with finite expected runtime. Most recently in 2018, Abe et al. \cite{abe} developed the first Las Vegas five-card AND protocol using only uniform closed shuffles by modifying the original five-card trick protocol. The important protocols developed so far are shown in Table \ref{upperbound}.

\begin{table}
	\centering
	\begin{tabular}{|c|c|c|c|c|c|}
		\hline
		\multirow{2}{*}{} & \multicolumn{2}{c|}{\textbf{Card}} & \multicolumn{3}{c|}{\textbf{Properties}} \\ \cline{2-6}
		& \textbf{\#colors} & \textbf{\#cards} & \textbf{finite} & \textbf{uniform} & \textbf{closed} \\ \hline \hline
		Cr\'{e}peau-Kilian \cite{crepeau}, 1993 & 4 & 10 & no & yes & yes \\ \hline
		Niemi-Renvall \cite{niemi}, 1998 & 2 & 12 & no & yes & yes \\ \hline
		Stiglic \cite{stiglic}, 2001 & 2 & 8 & no & yes & yes \\ \hline
		Mizuki-Sone \cite{mizuki09}, 2009 & 2 & 6 & yes & yes & yes \\ \hline
		Koch et al. \cite[\S4]{koch}, 2015 & 2 & 4 & no & no & yes \\ \hline
		Koch et al. \cite[\S5]{koch}, 2015 & 2 & 5 & yes & no & no \\ \hline
		Abe et al. \cite{abe}, 2018 & 2 & 5 & no & yes & yes \\ \hline \hline
		\textbf{Ours (\S2)} & 2 & 4 & no & yes & no \\ \hline
		\textbf{Ours (\S3)} & 2 & 5 & yes & yes & no \\ \hline
	\end{tabular}
	\caption{Previous development of committed-format AND protocols}
	\label{upperbound}
\end{table}

\subsection{Lower Bound}
On the other hand, several lower bounds of the minimum required number of two-color cards for an AND protocol subject to different restrictions have been proved. Koch et al. \cite[\S6]{koch} showed that there is no four-card AND protocol with finite runtime. Kastner et al. \cite{kastner} later proved that there is no finite five-card AND protocol using only closed shuffles, and no restart-free Las Vegas four-card AND protocol using only uniform closed shuffles.

Regarding the runtime, finiteness of shuffles, and closedness of shuffles, there are eight possible combinations of restrictions. The best lower bound and upper bound of the minimum required number of cards subject to each possible combination are shown in Table \ref{lowerbound}.

\begin{table}
	\centering
	\begin{tabular}{|c|c|c|c|c|}
		\hline
		\textbf{Runtime} & \textbf{Shuffle} & \textbf{\thead{Min.\\ \#Cards}} & \textbf{Lower Bound} & \textbf{Upper Bound} \\ \hline \hline
		\multirow{4}{*}{\thead{restart-free\\ Las Vegas}} & -- & 4 & \multirow{3}{*}{trivial} & \multirow{2}{*}{Koch et al. \cite[\S4]{koch}, 2015} \\ \cline{2-3}
		& closed & 4 & & \\ \cline{2-3} \cline{5-5}
		& uniform & 4 & & \textbf{Ours (\S2)} \\ \cline{2-5}
		& uniform closed & 5 & Kastner et al. \cite[\S7]{kastner}, 2017 & Abe et al. \cite{abe}, 2018 \\ \hline
		\multirow{4}{*}{finite} & -- & 5 & \multirow{2}{*}{Koch et al. \cite[\S6]{koch}, 2015} & Koch et al. \cite[\S5]{koch}, 2015 \\ \cline{2-3} \cline{5-5}
		& uniform & 5 & & \textbf{Ours (\S3)} \\ \cline{2-5}
		& closed & 6 & \multirow{2}{*}{Kastner et al. \cite[\S6]{kastner}, 2017} & \multirow{2}{*}{Mizuki-Sone \cite{mizuki09}, 2009} \\ \cline{2-3}
		& uniform closed & 6 & & \\ \hline
	\end{tabular}
	\caption{Minimum required number of two-color cards for a committed-format AND protocol, subject to each combination of runtime and shuffle restrictions}
	\label{lowerbound}
\end{table}

\subsection{Our Contribution}
Previously, the bounds in Table \ref{lowerbound} were all tight except in the third row (restart-free Las Vegas, uniform) where the trivial lower bound was four (since we need at least two cards for a commitment of each player's bit) but the upper bound was five (protocol of Abe et al. \cite{abe}), and the sixth row (finite, uniform) where the lower bound was five \cite{koch} but the upper bound was six (protocol of Mizuki and Sone \cite{mizuki09}).

In this paper, by modifying the protocols of Koch et al. \cite[\S4-5]{koch}, we introduce the first restart-free Las Vegas four-card AND protocol that uses only uniform shuffles, as well as the first finite five-card AND protocol that uses only uniform shuffles. This result also means that the lower bounds in the third and sixth rows of Table \ref{lowerbound} now become tight, thus completely answering the problem about the minimum required number of two-colored cards for a committed-format AND protocol subject to each combination of runtime and shuffle restrictions.

Shortly after this paper was first made public, Koch \cite[\S6]{koch2} also developed two protocols with the same properties as ours. This constitutes concurrent and independent work.

\section{Four-Card AND Protocol}
Starting at the four-card protocol of Koch et al. \cite[\S4]{koch}, we replace the closed but non-uniform shuffles by uniform but non-closed shuffles that have similar effects to the sequence. In this protocol, Alice's commitment and Bob's commitment are placed on the table in this order from left to right.

\subsection{Pseudocode}
A card is represented by a number based on its position on the table, with 1 being the leftmost card, 2 being the second card from the left, and so on. The following notions are also used in the pseudocode.
\begin{itemize}
	\item \textbf{(turn, \boldmath{$A$})} denotes flipping all cards in the set $A$.
	\item \textbf{visible} denotes a visible sequence of the cards from left to right, with ? being a face-down card.
	\item \textbf{(shuffle, \boldmath{$\Pi$})} denotes a uniform shuffle of the deck on the set $\Pi$ of permutations.
	\item \textbf{(perm, \boldmath{$\sigma$})} denotes rearranging the deck into a permutation $\sigma$.
	\item \textbf{(result, \boldmath{$x$}, \boldmath{$y$})} denotes outputting a commitment of card $x$ and card $y$, in this order.
\end{itemize}
\newpage

\begin{framed}
\noindent
\begin{tabbing}
(shuffle, \{id, (1 3)(2 4)\}) \\
(shuffle, \{id, (2 3)\}) \\
(turn, \{2\}) \\
\textbf{if} visible = (?,$\heartsuit$,?,?) \textbf{then} \\
\hspace{10mm} (turn, \{2\}) \\
\hspace{10mm} (shuffle, \{id, (3 4)\}) \\
\hspace{4.5mm} $\dagger$ \hspace{1mm} (shuffle, \{id, (3 4), (1 4 2 3)\}) \\
\hspace{10mm} (turn, \{4\}) \\
\hspace{10mm} \textbf{if} visible = (?,?,?,$\heartsuit$) \textbf{then} \\
\hspace{20mm} (\text{result}, 3, 2) \\
\hspace{10mm} \textbf{else} \\
\hspace{20mm} (turn, \{4\}) \\
\hspace{20mm} (shuffle, \{id, (1 2)\}) \\
\hspace{20mm} (perm, (2 3 4)) \\
\hspace{20mm} \textbf{goto $\star$} \\
\textbf{else} \\
\hspace{10mm} (turn, \{2\}) \\
\hspace{10mm} (shuffle, \{id, (1 3)\}) \\
\hspace{4.5mm} $\star$ \hspace{1mm} (shuffle, \{id, (1 3), (1 2 3 4)\}) \\
\hspace{10mm} (turn, \{1\}) \\
\hspace{10mm} \textbf{if} visible = ($\clubsuit$,?,?,?) \textbf{then} \\
\hspace{20mm} (result, 2, 3) \\
\hspace{10mm} \textbf{else} \\
\hspace{20mm} (turn, \{1\}) \\
\hspace{20mm} (shuffle, \{id, (2 4)\}) \\
\hspace{20mm} (perm, (1 2 3)) \\
\hspace{20mm} \textbf{goto $\dagger$}
\end{tabbing}
\end{framed}

\subsection{Proof of Correctness and Security}
We can easily verify the correctness of the protocol by keeping track of every possible sequence of the cards throughout the protocol. For the security, note that the shuffle and perm actions never reveal new information about the inputs; the only action that may reveal new information is the turn action. When we turn a set of cards face-up, we have to be sure that the probability to observe a visible sequence of cards is independent of the inputs $a$ and $b$.

In this paper we use a KWH-tree, a tool developed by Koch et al. \cite{koch}, to help verify the correctness and security of the protocol. $X_{00}$, $X_{01}$, $X_{10}$, and $X_{11}$ denote the probabilities of $(a,b)$ being $(0,0)$, $(0,1)$, $(1,0)$, and $(1,1)$, respectively, with shorthands $X_0 = X_{00}+X_{01}+X_{10}$ and $X_1 = X_{11}$ being used. Also, a polynomial denotes the conditional probability that the sequence of the cards is the one next to the polynomial, given the current view of the deck.

The KWH-tree of our four-card AND protocol is given in Figure 1. From the KWH-tree, We can verify that a correct commitment to $a \wedge b$ is obtained as a result, and that the sum of polynomials in every box equals to $X_0+X_1$, implying that no information about $a$ or $b$ leaks. This protocol is clearly restart-free. Also, at the final separating points (the boxes marked with an asterisk), the protocol terminates with probability $\frac{1}{3}$ and re-enter a branch on the other side with probability $\frac{2}{3}$. Therefore, the expected number of times it goes through the branches is $$\frac{1}{3}\left(1 + 2\left(\frac{2}{3}\right) + 3\left(\frac{2}{3}\right)^2 + ...\right) = 3,$$ thus having a finite expected runtime.

\section{Five-Card AND Protocol}
Starting at the five-card protocol of Koch et al. \cite[\S5]{koch}, we replace the closed but non-uniform shuffles by uniform but non-closed shuffles that have similar effects to the sequence. In this protocol, Alice's commitment, Bob's commitment, and an additional $\heartsuit$ card are placed on the table in this order from left to right.

\subsection{Pseudocode}
\begin{framed}
\noindent
\begin{tabbing}
(shuffle, \{id, (1 3)(2 4)\}) \\
(shuffle, \{id, (2 3)\}) \\
(turn, \{2\}) \\
\textbf{if} visible = (?,$\heartsuit$,?,?,?) \textbf{then} \\
\hspace{10mm} (turn, \{2\}) \\
\hspace{10mm} (shuffle, \{id, (3 4)\}) \\
\hspace{10mm} (shuffle, \{id, (3 4), (1 4 2 3)\}) \\
\hspace{10mm} (turn, \{4\}) \\
\hspace{10mm} \textbf{if} visible = (?,?,?,$\heartsuit$,?) \textbf{then} \\
\hspace{20mm} (\text{result}, 3, 2) \\
\hspace{10mm} \textbf{else} \\
\hspace{20mm} (turn, \{4\}) \\
\hspace{20mm} (shuffle, \{id, (1 2)\}) \\
\hspace{20mm} (perm, (2 3 4)) \\
\hspace{20mm} \textbf{goto $\star$} \\
\textbf{else} \\
\hspace{10mm} (turn, \{2\}) \\
\hspace{10mm} (shuffle, \{id, (1 3)\}) \\
\hspace{4.5mm} $\star$ \hspace{1mm} (shuffle, \{id, (1 3), (1 2)(3 5 4)\}) \\
\hspace{10mm} (turn, \{3\}) \\
\hspace{10mm} \textbf{if} visible = (?,?$\clubsuit$,?,?) \textbf{then} \\
\hspace{20mm} (result, 2, 1) \\
\hspace{10mm} \textbf{else} \\
\hspace{20mm} (result, 1, 4)
\end{tabbing}
\end{framed}

\subsection{Proof of Correctness and Security}
The KWH-tree of our five-card AND protocol is given in Figure 2. From the KWH-tree, We can verify that a correct commitment to $a \wedge b$ is obtained as a result, and that the sum of polynomials in every box equals to $X_0+X_1$, implying that no information about $a$ or $b$ leaks. This protocol clearly terminates in finite time since there is no cycle in the KWH-tree.

\section{Conclusion and Future Work}
In this paper, we introduce a restart-free Las Vegas four-card AND protocol and a finite five-card AND protocol, both using only uniform shuffles. This result also completely answers the problem about the minimum required number of two-colored cards for a committed-format AND protocol subject to each combination of runtime and shuffle restrictions.

The existing lower bounds, however, cover only the case with two-color cards. An interesting question is that whether the minimum required number of cards can be lowered if we allow more than two colors. For example, is there a finite five-card AND protocol using only closed shuffles if three-color cards are allowed?

\newpage

\appendix

\section*{KWH-Tree: Four-Card Protocol}
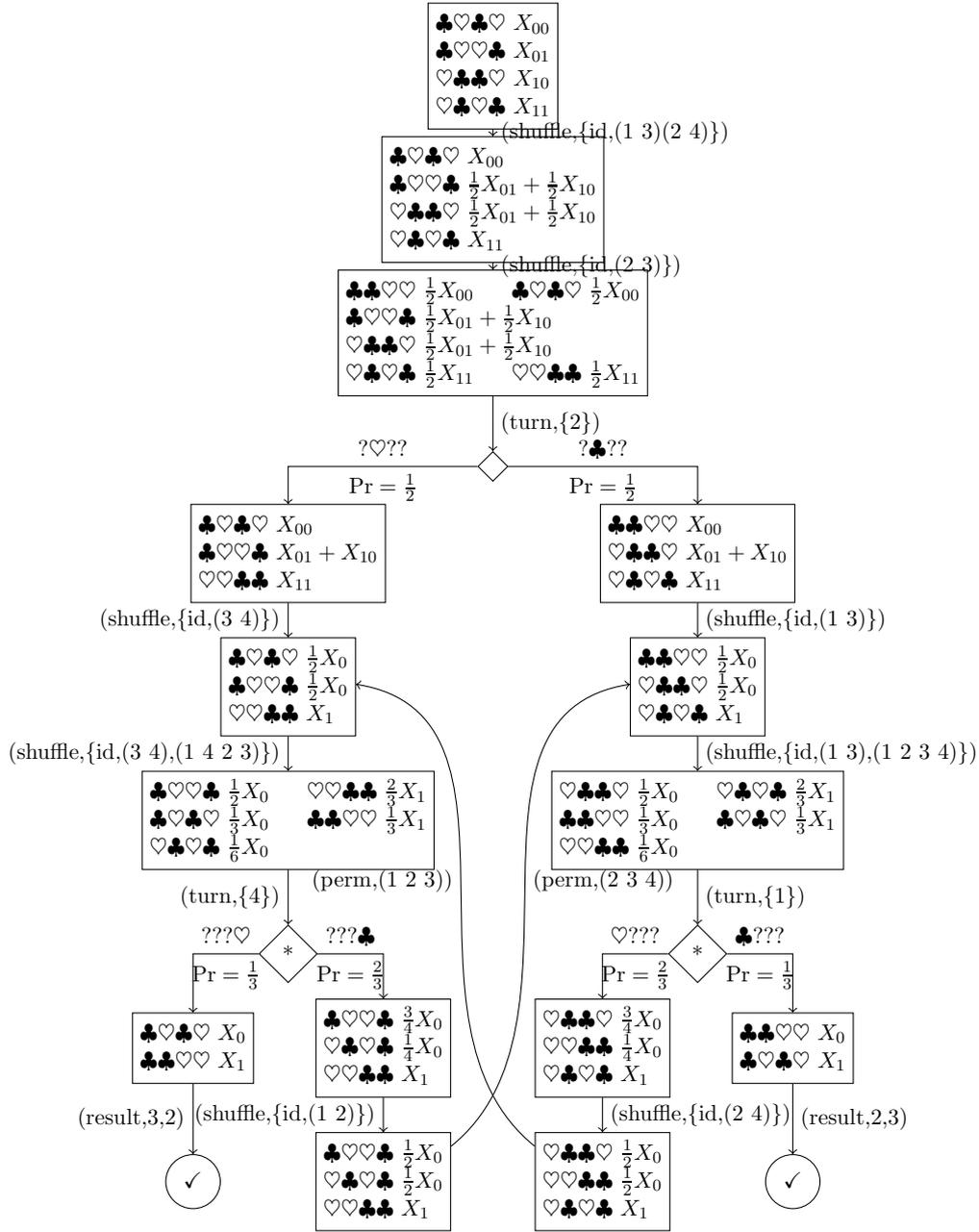
\begin{figure}[H]
\begin{tikzpicture}[node distance=2.3cm, auto, every node/.style={scale=0.8}]
    \node [block] (a1) {\makecell[l]{
    $\clubsuit$$\heartsuit$$\clubsuit$$\heartsuit$ $X_{00}$ \\
    $\clubsuit$$\heartsuit$$\heartsuit$$\clubsuit$ $X_{01}$ \\
    $\heartsuit$$\clubsuit$$\clubsuit$$\heartsuit$ $X_{10}$ \\
    $\heartsuit$$\clubsuit$$\heartsuit$$\clubsuit$ $X_{11}$
    }};
    \node [block, below of=a1] (a2) {\makecell[l]{
    $\clubsuit$$\heartsuit$$\clubsuit$$\heartsuit$ $X_{00}$ \\
    $\clubsuit$$\heartsuit$$\heartsuit$$\clubsuit$ $\frac{1}{2}X_{01} + \frac{1}{2}X_{10}$ \\
    $\heartsuit$$\clubsuit$$\clubsuit$$\heartsuit$ $\frac{1}{2}X_{01} + \frac{1}{2}X_{10}$ \\
    $\heartsuit$$\clubsuit$$\heartsuit$$\clubsuit$ $X_{11}$
    }};
    \node [block, below of=a2] (a3) {\makecell[l]{
    $\clubsuit$$\clubsuit$$\heartsuit$$\heartsuit$ $\frac{1}{2}X_{00}$ \hspace{1em} $\clubsuit$$\heartsuit$$\clubsuit$$\heartsuit$ $\frac{1}{2}X_{00}$ \\
    $\clubsuit$$\heartsuit$$\heartsuit$$\clubsuit$ $\frac{1}{2}X_{01} + \frac{1}{2}X_{10}$ \\
    $\heartsuit$$\clubsuit$$\clubsuit$$\heartsuit$ $\frac{1}{2}X_{01} + \frac{1}{2}X_{10}$ \\
    $\heartsuit$$\clubsuit$$\heartsuit$$\clubsuit$ $\frac{1}{2}X_{11}$ \hspace{1em} $\heartsuit$$\heartsuit$$\clubsuit$$\clubsuit$ $\frac{1}{2}X_{11}$
    }};
    \node [branch, below of=a3] (p1) {};
    \node [rectangle, left of=p1, node distance=3.5cm] (g1) {};
    \node [block, below of=g1, node distance=1.5cm] (b1) {\makecell[l]{
    $\clubsuit$$\heartsuit$$\clubsuit$$\heartsuit$ $X_{00}$ \\
    $\clubsuit$$\heartsuit$$\heartsuit$$\clubsuit$ $X_{01}+X_{10}$ \\
    $\heartsuit$$\heartsuit$$\clubsuit$$\clubsuit$ $X_{11}$
    }};
    \node [block, below of=b1] (b2) {\makecell[l]{
    $\clubsuit$$\heartsuit$$\clubsuit$$\heartsuit$ $\frac{1}{2}X_0$ \\
    $\clubsuit$$\heartsuit$$\heartsuit$$\clubsuit$ $\frac{1}{2}X_0$ \\
    $\heartsuit$$\heartsuit$$\clubsuit$$\clubsuit$ $X_1$
    }};
    \node [block, below of=b2] (b3) {\makecell[l]{
    $\clubsuit$$\heartsuit$$\heartsuit$$\clubsuit$ $\frac{1}{2}X_0$ \hspace{1em} $\heartsuit$$\heartsuit$$\clubsuit$$\clubsuit$ $\frac{2}{3}X_1$ \\
    $\clubsuit$$\heartsuit$$\clubsuit$$\heartsuit$ $\frac{1}{3}X_0$ \hspace{1em} $\clubsuit$$\clubsuit$$\heartsuit$$\heartsuit$ $\frac{1}{3}X_1$ \\
    $\heartsuit$$\clubsuit$$\heartsuit$$\clubsuit$ $\frac{1}{6}X_0$
    }};
    \node [branch, below of=b3] (p2) {*};
    \node [block, below left of=p2] (b4) {\makecell[l]{
    $\clubsuit$$\heartsuit$$\clubsuit$$\heartsuit$ $X_0$ \\
    $\clubsuit$$\clubsuit$$\heartsuit$$\heartsuit$ $X_1$
    }};
    \node [finish, below of=b4] (f1) {\makecell[l]{\checkmark}};
    \node [block, below right of=p2] (b5) {\makecell[l]{
    $\clubsuit$$\heartsuit$$\heartsuit$$\clubsuit$ $\frac{3}{4}X_0$ \\
    $\heartsuit$$\clubsuit$$\heartsuit$$\clubsuit$ $\frac{1}{4}X_0$ \\
    $\heartsuit$$\heartsuit$$\clubsuit$$\clubsuit$ $X_1$
    }};
    \node [block, below of=b5] (b6) {\makecell[l]{
    $\clubsuit$$\heartsuit$$\heartsuit$$\clubsuit$ $\frac{1}{2}X_0$ \\
    $\heartsuit$$\clubsuit$$\heartsuit$$\clubsuit$ $\frac{1}{2}X_0$ \\
    $\heartsuit$$\heartsuit$$\clubsuit$$\clubsuit$ $X_1$
    }};
    \node [rectangle, right of=p1, node distance=3.5cm] (g2) {};
    \node [block, below of=g2, node distance=1.5cm] (c1) {\makecell[l]{
    $\clubsuit$$\clubsuit$$\heartsuit$$\heartsuit$ $X_{00}$ \\
    $\heartsuit$$\clubsuit$$\clubsuit$$\heartsuit$ $X_{01}+X_{10}$ \\
    $\heartsuit$$\clubsuit$$\heartsuit$$\clubsuit$ $X_{11}$
    }};
    \node [block, below of=c1] (c2) {\makecell[l]{
    $\clubsuit$$\clubsuit$$\heartsuit$$\heartsuit$ $\frac{1}{2}X_0$ \\
    $\heartsuit$$\clubsuit$$\clubsuit$$\heartsuit$ $\frac{1}{2}X_0$ \\
    $\heartsuit$$\clubsuit$$\heartsuit$$\clubsuit$ $X_1$
    }};
    \node [block, below of=c2] (c3) {\makecell[l]{
    $\heartsuit$$\clubsuit$$\clubsuit$$\heartsuit$ $\frac{1}{2}X_0$ \hspace{1em} $\heartsuit$$\clubsuit$$\heartsuit$$\clubsuit$ $\frac{2}{3}X_1$ \\
    $\clubsuit$$\clubsuit$$\heartsuit$$\heartsuit$ $\frac{1}{3}X_0$ \hspace{1em} $\clubsuit$$\heartsuit$$\clubsuit$$\heartsuit$ $\frac{1}{3}X_1$ \\
    $\heartsuit$$\heartsuit$$\clubsuit$$\clubsuit$ $\frac{1}{6}X_0$
    }};
    \node [branch, below of=c3] (p3) {*};
    \node [block, below right of=p3] (c4) {\makecell[l]{
    $\clubsuit$$\clubsuit$$\heartsuit$$\heartsuit$ $X_0$ \\
    $\clubsuit$$\heartsuit$$\clubsuit$$\heartsuit$ $X_1$
    }};
    \node [finish, below of=c4] (f2) {\makecell[l]{\checkmark}};
    \node [block, below left of=p3] (c5) {\makecell[l]{
    $\heartsuit$$\clubsuit$$\clubsuit$$\heartsuit$ $\frac{3}{4}X_0$ \\
    $\heartsuit$$\heartsuit$$\clubsuit$$\clubsuit$ $\frac{1}{4}X_0$ \\
    $\heartsuit$$\clubsuit$$\heartsuit$$\clubsuit$ $X_1$
    }};
    \node [block, below of=c5] (c6) {\makecell[l]{
    $\heartsuit$$\clubsuit$$\clubsuit$$\heartsuit$ $\frac{1}{2}X_0$ \\
    $\heartsuit$$\heartsuit$$\clubsuit$$\clubsuit$ $\frac{1}{2}X_0$ \\
    $\heartsuit$$\clubsuit$$\heartsuit$$\clubsuit$ $X_1$
    }};
    
    \path [line] (a1) -- node{(shuffle,\{id,(1 3)(2 4)\})} (a2);
    \path [line] (a2) -- node{(shuffle,\{id,(2 3)\})} (a3);
    \path [line] (a3) -- node{(turn,\{2\})} (p1);
    \path [line] (p1) -| node[near start, above]{?$\heartsuit$??} node[near start, below]{$\Pr = \frac{1}{2}$} (b1);
    \path [line] (b1) -- node[left]{(shuffle,\{id,(3 4)\})} (b2);
    \path [line] (b2) -- node[left]{(shuffle,\{id,(3 4),(1 4 2 3)\})} (b3);
    \path [line] (b3) -- node[left]{(turn,\{4\})} (p2);
    \path [line] (p2) -| node[near start, above]{???$\heartsuit$} node[near start, below]{$\Pr = \frac{1}{3}$} (b4);
    \path [line] (b4) -- node[left]{(result,3,2)} (f1);
    \path [line] (p2) -| node[near start, above]{???$\clubsuit$} node[near start, below]{$\Pr = \frac{2}{3}$} (b5);
    \path [line] (b5) -- node[left]{(shuffle,\{id,(1 2)\})} (b6);
    \path [line, bend left=0] (b6) to [out=-30, in=120] node[right]{(perm,(2 3 4))} (c2);
    \path [line] (p1) -| node[near start, above]{?$\clubsuit$??} node[near start, below]{$\Pr = \frac{1}{2}$} (c1);
    \path [line] (c1) -- node{(shuffle,\{id,(1 3)\})} (c2);
    \path [line] (c2) -- node{(shuffle,\{id,(1 3),(1 2 3 4)\})} (c3);
    \path [line] (c3) -- node{(turn,\{1\})} (p3);
    \path [line] (p3) -| node[near start, above]{$\clubsuit$???} node[near start, below]{$\Pr = \frac{1}{3}$} (c4);
    \path [line] (c4) -- node{(result,2,3)} (f2);
    \path [line] (p3) -| node[near start, above]{$\heartsuit$???} node[near start, below]{$\Pr = \frac{2}{3}$} (c5);
    \path [line] (c5) -- node{(shuffle,\{id,(2 4)\})} (c6);
    \path [line, bend right=0] (c6) to [out=30, in=-120] node[left]{(perm,(1 2 3))} (b2);
\end{tikzpicture}

\begin{minipage}{\textwidth}
	\centering
	\caption{A KWH-tree of the four-card AND Protocol}
\end{minipage}
\end{figure}

\section*{KWH-Tree: Five-Card Protocol}
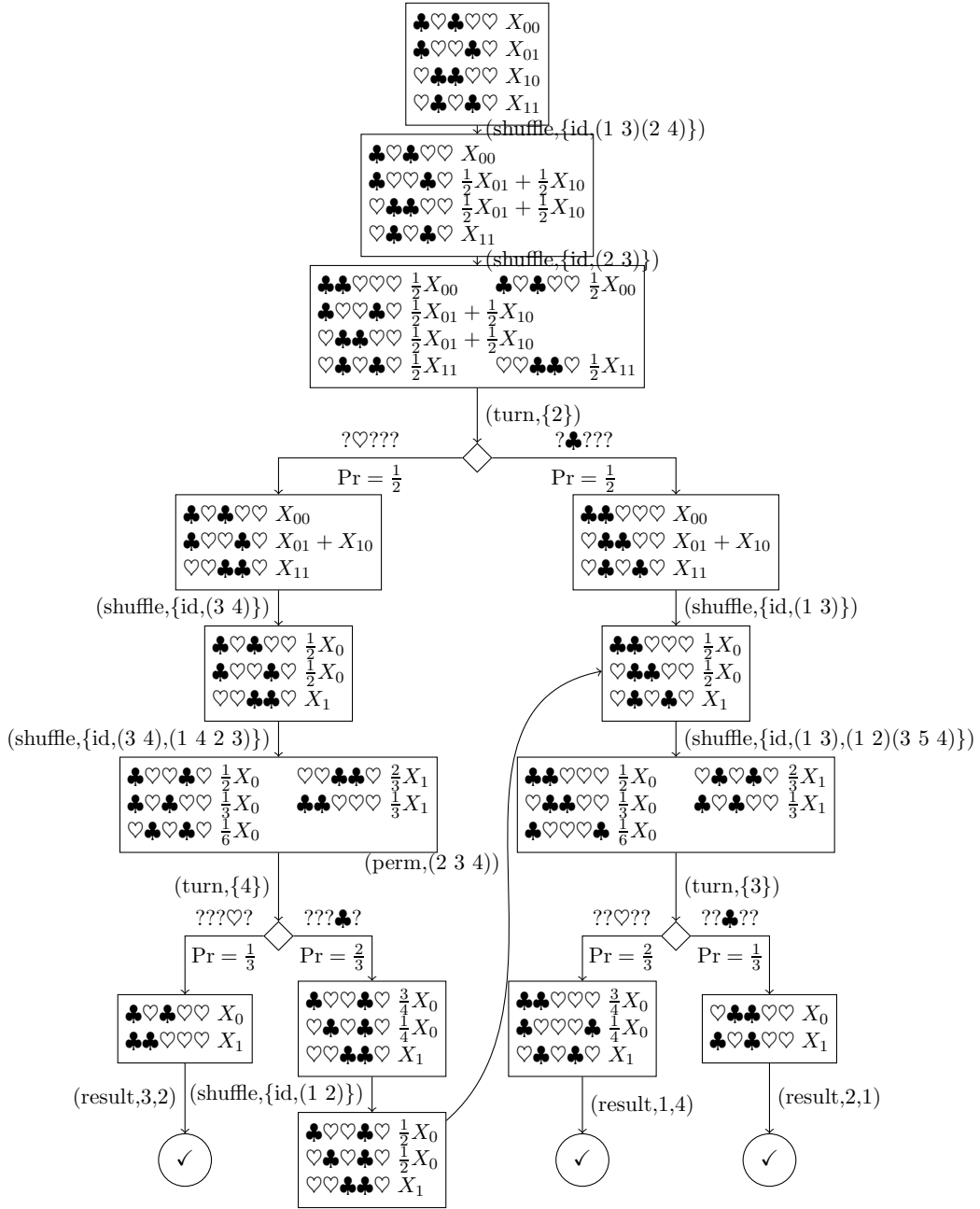
\begin{figure}[H]
\centering
\begin{tikzpicture}[node distance=2.33cm, auto, every node/.style={scale=0.8}]
    \node [block] (a1) {\makecell[l]{
    $\clubsuit$$\heartsuit$$\clubsuit$$\heartsuit$$\heartsuit$ $X_{00}$ \\
    $\clubsuit$$\heartsuit$$\heartsuit$$\clubsuit$$\heartsuit$ $X_{01}$ \\
    $\heartsuit$$\clubsuit$$\clubsuit$$\heartsuit$$\heartsuit$ $X_{10}$ \\
    $\heartsuit$$\clubsuit$$\heartsuit$$\clubsuit$$\heartsuit$ $X_{11}$
    }};
    \node [block, below of=a1] (a2) {\makecell[l]{
    $\clubsuit$$\heartsuit$$\clubsuit$$\heartsuit$$\heartsuit$ $X_{00}$ \\
    $\clubsuit$$\heartsuit$$\heartsuit$$\clubsuit$$\heartsuit$ $\frac{1}{2}X_{01} + \frac{1}{2}X_{10}$ \\
    $\heartsuit$$\clubsuit$$\clubsuit$$\heartsuit$$\heartsuit$ $\frac{1}{2}X_{01} + \frac{1}{2}X_{10}$ \\
    $\heartsuit$$\clubsuit$$\heartsuit$$\clubsuit$$\heartsuit$ $X_{11}$
    }};
    \node [block, below of=a2] (a3) {\makecell[l]{
    $\clubsuit$$\clubsuit$$\heartsuit$$\heartsuit$$\heartsuit$ $\frac{1}{2}X_{00}$ \hspace{1em} $\clubsuit$$\heartsuit$$\clubsuit$$\heartsuit$$\heartsuit$ $\frac{1}{2}X_{00}$ \\
    $\clubsuit$$\heartsuit$$\heartsuit$$\clubsuit$$\heartsuit$ $\frac{1}{2}X_{01} + \frac{1}{2}X_{10}$ \\
    $\heartsuit$$\clubsuit$$\clubsuit$$\heartsuit$$\heartsuit$ $\frac{1}{2}X_{01} + \frac{1}{2}X_{10}$ \\
    $\heartsuit$$\clubsuit$$\heartsuit$$\clubsuit$$\heartsuit$ $\frac{1}{2}X_{11}$ \hspace{1em} $\heartsuit$$\heartsuit$$\clubsuit$$\clubsuit$$\heartsuit$ $\frac{1}{2}X_{11}$
    }};
    \node [branch, below of=a3] (p1) {};
    \node [rectangle, left of=p1, node distance=3.5cm] (g1) {};
    \node [block, below of=g1, node distance=1.5cm] (b1) {\makecell[l]{
    $\clubsuit$$\heartsuit$$\clubsuit$$\heartsuit$$\heartsuit$ $X_{00}$ \\
    $\clubsuit$$\heartsuit$$\heartsuit$$\clubsuit$$\heartsuit$ $X_{01}+X_{10}$ \\
    $\heartsuit$$\heartsuit$$\clubsuit$$\clubsuit$$\heartsuit$ $X_{11}$
    }};
    \node [block, below of=b1] (b2) {\makecell[l]{
    $\clubsuit$$\heartsuit$$\clubsuit$$\heartsuit$$\heartsuit$ $\frac{1}{2}X_0$ \\
    $\clubsuit$$\heartsuit$$\heartsuit$$\clubsuit$$\heartsuit$ $\frac{1}{2}X_0$ \\
    $\heartsuit$$\heartsuit$$\clubsuit$$\clubsuit$$\heartsuit$ $X_1$
    }};
    \node [block, below of=b2] (b3) {\makecell[l]{
    $\clubsuit$$\heartsuit$$\heartsuit$$\clubsuit$$\heartsuit$ $\frac{1}{2}X_0$ \hspace{1em} $\heartsuit$$\heartsuit$$\clubsuit$$\clubsuit$$\heartsuit$ $\frac{2}{3}X_1$ \\
    $\clubsuit$$\heartsuit$$\clubsuit$$\heartsuit$$\heartsuit$ $\frac{1}{3}X_0$ \hspace{1em} $\clubsuit$$\clubsuit$$\heartsuit$$\heartsuit$$\heartsuit$ $\frac{1}{3}X_1$ \\
    $\heartsuit$$\clubsuit$$\heartsuit$$\clubsuit$$\heartsuit$ $\frac{1}{6}X_0$
    }};
    \node [branch, below of=b3] (p2) {};
    \node [block, below left of=p2] (b4) {\makecell[l]{
    $\clubsuit$$\heartsuit$$\clubsuit$$\heartsuit$$\heartsuit$ $X_0$ \\
    $\clubsuit$$\clubsuit$$\heartsuit$$\heartsuit$$\heartsuit$ $X_1$
    }};
    \node [finish, below of=b4] (f1) {\makecell[l]{\checkmark}};
    \node [block, below right of=p2] (b5) {\makecell[l]{
    $\clubsuit$$\heartsuit$$\heartsuit$$\clubsuit$$\heartsuit$ $\frac{3}{4}X_0$ \\
    $\heartsuit$$\clubsuit$$\heartsuit$$\clubsuit$$\heartsuit$ $\frac{1}{4}X_0$ \\
    $\heartsuit$$\heartsuit$$\clubsuit$$\clubsuit$$\heartsuit$ $X_1$
    }};
    \node [block, below of=b5] (b6) {\makecell[l]{
    $\clubsuit$$\heartsuit$$\heartsuit$$\clubsuit$$\heartsuit$ $\frac{1}{2}X_0$ \\
    $\heartsuit$$\clubsuit$$\heartsuit$$\clubsuit$$\heartsuit$ $\frac{1}{2}X_0$ \\
    $\heartsuit$$\heartsuit$$\clubsuit$$\clubsuit$$\heartsuit$ $X_1$
    }};
    \node [rectangle, right of=p1, node distance=3.5cm] (g2) {};
    \node [block, below of=g2, node distance=1.5cm] (c1) {\makecell[l]{
    $\clubsuit$$\clubsuit$$\heartsuit$$\heartsuit$$\heartsuit$ $X_{00}$ \\
    $\heartsuit$$\clubsuit$$\clubsuit$$\heartsuit$$\heartsuit$ $X_{01}+X_{10}$ \\
    $\heartsuit$$\clubsuit$$\heartsuit$$\clubsuit$$\heartsuit$ $X_{11}$
    }};
    \node [block, below of=c1] (c2) {\makecell[l]{
    $\clubsuit$$\clubsuit$$\heartsuit$$\heartsuit$$\heartsuit$ $\frac{1}{2}X_0$ \\
    $\heartsuit$$\clubsuit$$\clubsuit$$\heartsuit$$\heartsuit$ $\frac{1}{2}X_0$ \\
    $\heartsuit$$\clubsuit$$\heartsuit$$\clubsuit$$\heartsuit$ $X_1$
    }};
    \node [block, below of=c2] (c3) {\makecell[l]{
    $\clubsuit$$\clubsuit$$\heartsuit$$\heartsuit$$\heartsuit$ $\frac{1}{2}X_0$ \hspace{1em} $\heartsuit$$\clubsuit$$\heartsuit$$\clubsuit$$\heartsuit$ $\frac{2}{3}X_1$ \\
    $\heartsuit$$\clubsuit$$\clubsuit$$\heartsuit$$\heartsuit$ $\frac{1}{3}X_0$ \hspace{1em} $\clubsuit$$\heartsuit$$\clubsuit$$\heartsuit$$\heartsuit$ $\frac{1}{3}X_1$ \\
    $\clubsuit$$\heartsuit$$\heartsuit$$\heartsuit$$\clubsuit$ $\frac{1}{6}X_0$
    }};
    \node [branch, below of=c3] (p3) {};
    \node [block, below left of=p3] (c4) {\makecell[l]{
    $\clubsuit$$\clubsuit$$\heartsuit$$\heartsuit$$\heartsuit$ $\frac{3}{4}X_0$ \\
    $\clubsuit$$\heartsuit$$\heartsuit$$\heartsuit$$\clubsuit$ $\frac{1}{4}X_0$ \\
    $\heartsuit$$\clubsuit$$\heartsuit$$\clubsuit$$\heartsuit$ $X_1$
    }};
    \node [finish, below of=c4] (f2) {\makecell[l]{\checkmark}};
    \node [block, below right of=p3] (c5) {\makecell[l]{
    $\heartsuit$$\clubsuit$$\clubsuit$$\heartsuit$$\heartsuit$ $X_0$ \\
    $\clubsuit$$\heartsuit$$\clubsuit$$\heartsuit$$\heartsuit$ $X_1$
    }};
    \node [finish, below of=c5] (f3) {\makecell[l]{\checkmark}};
    
    \path [line] (a1) -- node{(shuffle,\{id,(1 3)(2 4)\})} (a2);
    \path [line] (a2) -- node{(shuffle,\{id,(2 3)\})} (a3);
    \path [line] (a3) -- node{(turn,\{2\})} (p1);
    \path [line] (p1) -| node[near start, above]{?$\heartsuit$???} node[near start, below]{$\Pr = \frac{1}{2}$} (b1);
    \path [line] (b1) -- node[left]{(shuffle,\{id,(3 4)\})} (b2);
    \path [line] (b2) -- node[left]{(shuffle,\{id,(3 4),(1 4 2 3)\})} (b3);
    \path [line] (b3) -- node[left]{(turn,\{4\})} (p2);
    \path [line] (p2) -| node[near start, above]{???$\heartsuit$?} node[near start, below]{$\Pr = \frac{1}{3}$} (b4);
    \path [line] (b4) -- node[left]{(result,3,2)} (f1);
    \path [line] (p2) -| node[near start, above]{???$\clubsuit$?} node[near start, below]{$\Pr = \frac{2}{3}$} (b5);
    \path [line] (b5) -- node[left]{(shuffle,\{id,(1 2)\})} (b6);
    \path [line, bend left=0] (b6) to [out=-30, in=120] node[left]{(perm,(2 3 4))} (c2);
    \path [line] (p1) -| node[near start, above]{?$\clubsuit$???} node[near start, below]{$\Pr = \frac{1}{2}$} (c1);
    \path [line] (c1) -- node{(shuffle,\{id,(1 3)\})} (c2);
    \path [line] (c2) -- node{(shuffle,\{id,(1 3),(1 2)(3 5 4)\})} (c3);
    \path [line] (c3) -- node{(turn,\{3\})} (p3);
    \path [line] (p3) -| node[near start, above]{??$\heartsuit$??} node[near start, below]{$\Pr = \frac{2}{3}$} (c4);
    \path [line] (c4) -- node{(result,1,4)} (f2);
    \path [line] (p3) -| node[near start, above]{??$\clubsuit$??} node[near start, below]{$\Pr = \frac{1}{3}$} (c5);
    \path [line] (c5) -- node{(result,2,1)} (f3);
\end{tikzpicture}

\begin{minipage}{\textwidth}
	\centering
	\caption{A KWH-tree of the five-card AND Protocol}
\end{minipage}
\end{figure}

\end{document}